\documentclass[aip,jcp,graphicx]{revtex4-1}

\usepackage{graphicx}

\draft 

\begin{document}


\title{Anisotropic Dielectric Relaxation in Single Crystal H$_{2}$O Ice Ih from 80--250~K} 



\author{David T. W. Buckingham}
\author{V. Hugo Schmidt}
\author{John J. Neumeier}
\affiliation{Department of Physics, Montana State University, P.O. Box 173840, Bozeman, Montana, USA 59717-3840}


\date{\today}

\begin{abstract}
Three properties of the dielectric relaxation in ultra-pure single crystalline H$_{2}$O ice Ih were probed at temperatures between 80--250~K; the thermally stimulated depolarization current, static electrical conductivity, and dielectric relaxation time. The measurements were made with a guarded parallel-plate capacitor constructed of fused quartz with Au electrodes. The data agree with relaxation-based models and provide for the determination of activation energies, which suggest that relaxation in ice is dominated by Bjerrum defects below 140~K. Furthermore, anisotropy in the dielectric relaxation data reveals that molecular reorientations along the crystallographic \(c\)-axis are energetically favored over those along the \(a\)-axis between 80--140~K. These results lend support for the postulate of a shared origin between the dielectric relaxation dynamics and the thermodynamic partial proton-ordering in ice near 100~K, and suggest a preference for ordering along the \(c\)-axis.
\end{abstract}


\maketitle 

\section{Introduction}
\label{intro}

\begin{figure}[b!]
    \includegraphics[width=0.25\textwidth]{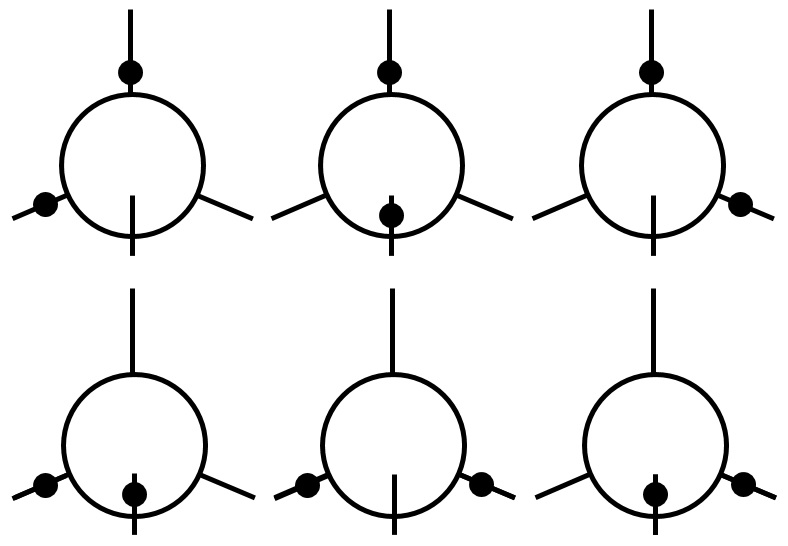}
		\caption{The six possible configurations of the two covalently bonded hydrogens (filled circles) around an oxygen (open circles) in the structure of ice Ih.}
		\label{H_config}
\end{figure}

H$_{2}$O ice is one of the most recognizable and highly-studied solids in the world.~\cite{petrenko} Still, some of its fundamental properties have mystified scientists since the early 20$^{\textrm{th}}$ century.~\cite{hobbs, debenedetti2003, ball2008, amann-winkel2016} Of the eighteen or so~\cite{bartels-rausch2012, debenedetti} different phases of crystalline ice, the most abundant and familiar is ice Ih. It has a hexagonal crystal structure~\cite{kuhs1986} defined by the ordered tetrahedral arrangement of the oxygen atoms. Within the limits of the ice rules,~\cite{bernal1933} however, the positions of the hydrogen atoms (or protons) are highly disordered,~\cite{pauling1935, haida1974, shi2013} with one of the six possible configurations seen in Figure~\ref{H_config}
around each oxygen atom. Ice XI~\cite{tajima1982} is the proton-ordered phase of hexagonal ice, with periodically repeating hydrogen configurations around the oxygen atoms. This order, paired with the fact that each free H$_{2}$O molecule has a dipole moment of magnitude~\cite{clough1973} \(6.186(1) \times 10^{-30}~\mathrm{C{\cdotp}m}\) in the direction parallel to the bisector of the H---O---H angle, causes ice XI to be ferroelectric along the \(c\)-axis.~\cite{jackson1995} But, because the protons in ice Ih are disordered, the bulk crystals have zero net dipole moment~\cite{pamuk2015} and are, therefore, paraelectric. It is possible to induce a polarization \(\vec{P}\) in ice~\cite{zaretskii1987} with the application of an external electric field \(\vec{E}\). By measuring changes in \(\vec{P}\) with time and temperature, properties related to dielectric relaxation can be determined.

\begin{figure}[t!]
    \includegraphics[width=0.32\textwidth]{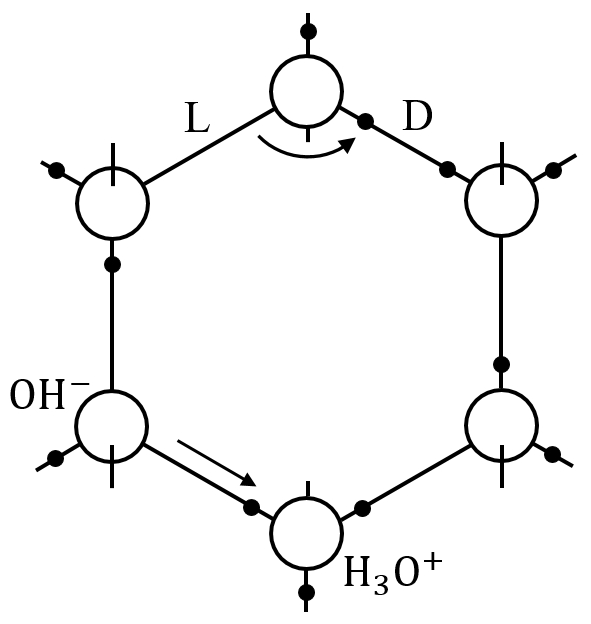}
		\caption{A 2-dimensional projection of the ice Ih crystal lattice demonstrating both the creation of Bjerrum defects, L and D, after a molecular rotation and the creation of ionic defects, OH$^{-}$ and H$_{3}$O$^{+}$, resulting from a proton jump along the bond. Open circles represent oxygen atoms and filled circles denote hydrogen. The O---O, O---H, and H---H distances are not drawn to scale.}
		\label{defects}
\end{figure}

Polarization in dielectric media is proportional to the applied field according to the electrostatic equation for a linear dielectric~\cite{griffiths}
\begin{equation}
\label{eq:polarization1}
\vec{P}=\epsilon_{0} \chi_{e} \vec{E},
\end{equation}
where \(\epsilon_{0}\) is the permittivity of free space and \(\chi_{e}\) is the electric susceptibility of the dielectric. In pure ice, this induced polarization is due to a partial ordering of the proton configuration, mainly by means of the orientational point defects commonly referred to as Bjerrum L- and D-defects.~\cite{bjerrum1952, jaccard1959, petrenko, zaretskii1987} An L-defect is a hydrogen bond between oxygen atoms that is left vacant (i.e. without a hydrogen atom) after a molecular rotation. Similarly, a D-defect is a bond which is left doubly occupied by two protons. Ionic point defects H$_{3}$O$^{+}$ and OH$^{-}$ can also contribute to polarization and conduction in ice Ih, but they tend to dominate in ice that is heavily-doped with impurities like KOH.~\cite{kawada1997} These four point defects are shown in Figure~\ref{defects}.

This manuscript outlines measurements of the bulk dielectric polarization in ice Ih (simply referred to as ice from here) as a function of time and temperature. The results are used to determine trends in the thermally stimulated depolarization (TSD) current, and values of the static conductivity and dielectric relaxation time, between 80--250~K. They are in close agreement with those determined by prior researchers,~\cite{johari1975, bullemer1969, sasaki2016} and reveal anisotropy in the dielectric relaxation dynamics of ice along the crystallographic \(a\)- and \(c\)-axes (simply referred to as \(a\) and \(c\), respectively, from here) that suggests a preference for molecular reorientations along \(c\).

\section{Theoretical Background}
\label{theory}

The theories of the TSD currents~\cite{johari1975, loria1978, zaretskii1987, apekis1987}, static conductivity~\cite{jaccard1965, petrenko1993, hobbs, petrenko}, and dielectric relaxation time~\cite{kauzmann1942, eyring1935, johari1975, loria1978, zaretskii1991} for ice are well documented and have shown conformity with experiment. In this work, only a fraction of the complete theoretical knowledge on ice is required for the data analysis. The reader is thus referred to the above references for a complete description of the theories. Below is an outline of the theoretical background required for this work.

\subsection{Static conductivity \(\sigma_{s}\)}
\label{ssec:TSDtheory}

Electrical conductivity requires drift of carriers from unit cell to unit cell in response to an applied electric field. In contrast, polarization describes the dielectric motion that is confined within a unit cell. In ice, the response to an electric field consists of the motion of Bjerrum and ionic defects as mentioned in Section~\ref{intro}. Bjerrum defects move by means of successive rotations of water molecules, which can alter the polarization but cannot move charges (protons) from one molecule to another. Ionic defects move by means of successive motions of protons along hydrogen bonds, which also can alter the polarization but cannot move protons away from their original bonds. Accordingly, both Bjerrum and ionic defect motions are needed for conductivity to occur in ice.~\cite{petrenko} The activation energy for the conductivity will be the activation energy for the process which has the higher barrier, which for ice is ionic motion.

\begin{figure}[t!]
    \includegraphics[width=0.875\textwidth]{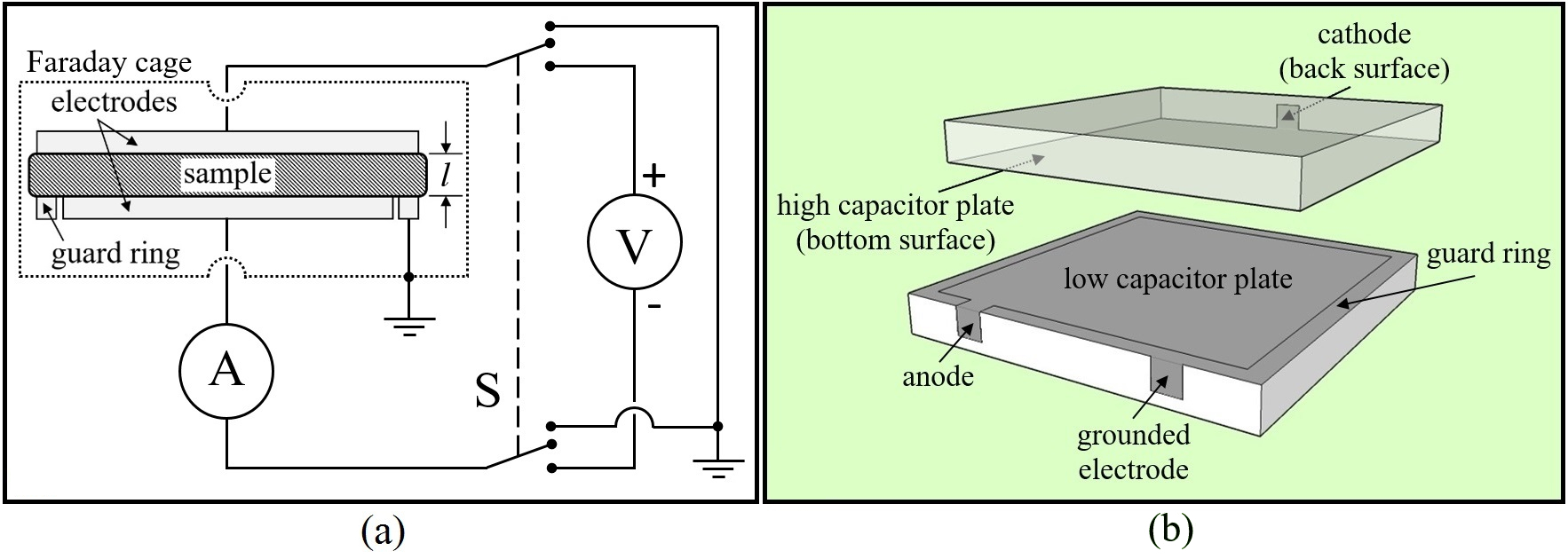}
		\caption{(a) The circuit diagram of the experiment, with a 2-D schematic of the dielectric cell and a sample of thickness \(l\) situated within a Faraday cage that shares a common ground with the guard ring. Components A, S, and V are an ammeter, a DPDT single-break level-switch, and a DC floating-voltage source, respectively. (b) An exploded view of the fused quartz dielectric cell, showing the high- and low-capacitor plates, the guard ring, and the electrodes made of Au.}
		\label{circuit_cell}
\end{figure}

The conduction process in ice can be partially modeled with two independent relations for the \textit{static conductivity} \(\sigma_{s}\). The first is the standard equation for stable ohmic conduction,
\begin{equation}
\label{eq:conductivity}
\sigma_{s} = \frac{I_{\sigma}}{V}\frac{l}{A}.
\end{equation}
The variables in Equation~\ref{eq:conductivity} are defined in reference to the diagrams shown in Figures~\ref{circuit_cell}(a) and \ref{circuit_cell}(b),
where \(A\) is the area of the low capacitor plate, \(l\) is the sample thickness, \(V\) is the applied voltage, and \(I_{\sigma}\) is the conduction current through the sample. The second relation for \(\sigma_{s}\) comes from the equation for an ionic semiconductor,
\begin{equation}
\label{eq:hugo1}
\sigma_{s} = n q \mu = \sigma_{0} \mathrm{e}^{-E_{\sigma}/k_{B}T},
\end{equation}
where both the carrier concentration \(n\) and the mobility \(\mu\) are thermally activated, \(q\) for ice is the proton charge \(e\), \(\sigma_{0}\) is the infinite temperature conductivity, \(E_{\sigma}\) is the \textit{activation energy of static conductivity}, \(k_{B}\) is Boltzmann's constant, and \(T\) is the temperature. The value of \(\sigma_{0}\) is determined by extrapolating the usually straight line in the plot of \(\log \sigma_{s}\) versus \(1/T\). The importance of \(\sigma_{0}\) lies in the fact that its magnitude can suggest whether the conduction process is intrinsic or extrinsic. Kittel~\cite{kittel} gives the expression
\begin{equation}
\label{eq:hugo2}
\sigma_{0} = \frac{N_{0} e^{2} p \nu a^{2}}{k_{B} T},
\end{equation}
where, for ice, \(N_{0}\) is the proton concentration, \(p\) is the probability that the carrier is an ionic defect (\(p \approx 1\) as \(T \rightarrow \infty\)), \(\nu \approx k_{B} T / h\) is the jump attempt frequency below the Debye temperature, \(h\) is Planck's constant, and \(a\) is the carrier jump distance along \(\vec{E}\). For a simple cubic crystal, \(N_{0} = 1 / a^{3}\), so in a rough but simple approximation, \(\sigma_{0} \approx e^{2} / h a\) if the conductivity is intrinsic.  Choosing \(a=2\)~$\mathrm{\AA}$, nearly the distance between two adjacent O--H$\cdots$O bonds, \(\sigma_{0} \approx 2 \times 10^{3}\)~S/cm.

If the conductivity is extrinsic, for instance by doping with a fraction \(f\) of KOH, we expect a fixed concentration \(f N_{0}\) of OH$^{-}$ ionic carriers and almost no H$_{3}$O$^{+}$ ionic carriers. Then, \(E_{\sigma}\) will be only the mobility activation energy, and the linear extrapolation of the semi-\(\log\) conductivity plot should give a \(\sigma_{0}\) which is a factor \(f\) lower than for the intrinsic \(\sigma_{0}\) value. If the crystal is not intentionally doped, \(f\) should be the level of unintentional doping. Sometimes the semi-\(\log\) plot will display a kink, going from a lower slope corresponding to extrinsic conductivity at lower temperatures to a higher slope corresponding to intrinsic conductivity at higher temperatures. It follows that at a given \(T\), extrinsic conductivity should never be lower than the intrinsic conductivity. Finally, the two types of ionic carriers can be expected to have different mobility activation energies, with the H$_{3}$O$^{+}$ energy being lower because its third proton is less energetically bound to its oxygen than the proton which would jump to the OH$^{-}$ ion in the \({\text{H---O---H}+\text{O---H}^{-} \rightarrow \text{H---O}^{-}+\text{H---O---H}}\) conduction process.

\subsection{Dielectric relaxation time \(\tau_{D}\)}
\label{ssec:tautheory}

The \textit{dielectric relaxation time} \(\tau_{D}\) is a measure of the time taken to polarize a dielectric following the application of an external electric field. This delay in the polarization can be modeled by the Debye relaxation process~\cite{debye} as extended by Petrenko and Whitworth,~\cite{petrenko} given by
\begin{equation}
\label{eq:rate}
\frac{dP}{dt}=\frac{1}{\tau_{D}}\left(P_{s}-P\right),
\end{equation}
where \(P=|\vec{P}|\) and \(P_{s}=\epsilon_{0} \chi_{s} |\vec{E}|\) is the static equilibrium polarization of the dielectric. For ice, the dielectric relaxation time depends on the temperature and purity of the sample.~\cite{petrenko, suga2005} 

The Debye relaxation time has been theoretically evaluated and extended by Kauzmann~\cite{kauzmann1942} from Eyring's rate-process theory~\cite{eyring1935} and applied to ice by Johari and Jones~\cite{johari1975} and Loria, \textit{et~al}.~\cite{loria1978} The general form of the dielectric relaxation time is given by
\begin{equation}
\label{eq:relaxation}
\tau_{D}\left(T\right)=\tau_{0}e^{E_{\tau}/k_{B}T},
\end{equation}
where \(\tau_{0}\) is a scaling factor and \(E_{\tau}\) is the \textit{activation energy of dielectric relaxation}. For the purposes of this manuscript \(\tau_{0}\), \(E_{\tau}\), and \(k_{B}\) can all be treated as constants, to first-order.~\cite{loria1978} 

Because of its reliability in measuring \(\tau_{D}\) on the order of hours or days, the \textit{voltage-step technique}~\cite{johari1975, zaretskii1991} was used to determine the \(\tau_{D}\) of ice at low \(T\). The theory of this technique is explained as follows. By `sandwiching' a sample between two ideal ohmic parallel-plate electrodes (see Figure~\ref{circuit_cell}(a)), applying a DC voltage across the electrodes, and measuring the current to the capacitor plates as a function of time, one can calculate \(\tau_{D}\) using the well-known theories for the \textit{displacement current},~\cite{maxwell} \(I_{d}\), and the \textit{proton conduction current},~\cite{jaccard1965} \(I_{\sigma}\). The displacement current is given by the dynamic equation
\begin{equation}
\label{eq:Idsp}
I_{d}(t)=\epsilon_{0}A\frac{d E}{d t}+A\frac{d P}{d t}.
\end{equation}
Since \(\vec{E}\) is static, the first term in Equation~\ref{eq:Idsp} is zero and, after combining with the solution to Equation~\ref{eq:rate},
\begin{equation}
\label{eq:polarization2}
P(t)=P_{s}\left(1-e^{-t/\tau_{D}}\right),
\end{equation}
\(I_{d}\) becomes
\begin{equation}
\label{eq:DispCurrent}
I_{d}(t)=\frac{A P_{s}}{\tau_{D}} e^{-t/\tau_{D}}.
\end{equation}
In the ohmic conduction process, the conduction current \(I_{\sigma}\) is related to the steady-state conductivity \(\sigma_{s}\) according to Equation~\ref{eq:conductivity}. This current is isothermally static and can therefore be treated as a constant. The sum of these currents gives an expression for the measured current across the capacitor as
\begin{equation}
\label{eq:current}
I(t)=I_{\sigma}+\frac{A P_{s}}{\tau_{D}} e^{-t/\tau_{D}}.
\end{equation}
It should be noted that this relation holds only if the system has one relaxation time. In some instances pure ice has been observed to exhibit more than one relaxation time.~\cite{zaretskii1991, apekis1987} However, in this work Equation~\ref{eq:current} provides good agreement with experiment. 

\section{Experiment}
\label{exp}

The water used to grow the ice single crystals was purified with reverse osmosis followed by filtration through a Milli-Q Advantage A10 System with a 0.22~$\mu$m Millipak polisher to obtain ultra-pure water~\cite{riche2006} with an electrical resistivity of 18.18~M$\Omega$~cm at 25 $^{\circ}$C, a total organic carbon concentration of 121~ppb, and a pH of 6.998. The water was degassed using a freeze-pump-thaw cycle~\cite{shriver} until gas evolution was no longer observed. A cylindrical single crystal ingot, of length 16~cm and diameter 2.5~cm, was formed using a zone-refining method similar to that of Bilgram, \textit{et~al}.~\cite{bilgram1973} In a cold room at 265~K, the samples were cut from the ingot and oriented using the optical polarization technique~\cite{hobbs} and a two-axis manual goniometer. The crystals were polished into square prisms of thickness \(l\), with the square faces being normal ($\pm$1.5$^{\circ}$) to \(a\) and \(c\). Because aging can affect some of the properties of ice,~\cite{petrenko} it is important to note that the sample was grown 10 months prior to measurement, annealed at 255~K for 7 months until it was oriented and annealed again for 3 months.  Between 1--12 h before each measurement, a sample with \(l\) typically between 5--7~mm was `sandwiched' between the capacitor plates of the dielectric cell, which was held together with two BeCu C-clamp springs.

An exploded view of the dielectric cell is shown in Figure~\ref{circuit_cell}(b). Its construction was similar to that outlined by Neumeier, \textit{et~al}.~\cite{neumeier2008} for their quartz dilatometer cell. It consists of two fused quartz square prisms of dimensions 2$\times$2$\times$0.25~cm$^{3}$. On one square face and one side of each platform, a 100~$\mathrm{\AA}$/1000~$\mathrm{\AA}$ Cr/Au film was vapor-deposited to form the conductive capacitor plates and electrode tabs, respectively. A 18~$\mu$m-thick line of gold, inset $\sim$1~mm from the edge of the low capacitor plate, was removed to create the guard ring shown in Figure~\ref{circuit_cell}(b). The electrical connections to the cell were made with annealed 25~$\mu$m-diameter platinum wire that was adhered to the electrodes with silver paint. The high- and low-capacitor plates were connected to the central conductors of coaxial cables. The cable shielding and guard ring were grounded to a Faraday cage that contained the cell and eliminated external electromagnetic fields. A Keithley Model 5517A Electrometer, capable of reliably measuring currents as low as 0.01~pA, was used for the current measurements. The circuit is shown in Figure~\ref{circuit_cell}(a).

The Faraday cage was mounted to a stainless-steel tube with an O-ring flange on top that suspended the cell within a hermetically-sealed cryostat, described in detail elsewhere.~\cite{neumeier2008} Before each measurement the Faraday-cage assembly (without the sample-cell assembly) was placed in the cryostat. To keep it free of condensed gases on cooling, the sample space was pumped to 10$^{-5}$~mbar at 350~K for 12 h with three intermediate flushes with pure He gas. The cryostat was then cooled with liquid nitrogen (LN$_{2}$) and held at 265~K until thermal equilibrium was reached, after which the cryostat was flooded with He and the assembly was removed. The cryostat was immediately capped and the assembly was quickly placed in an atmosphere of LN$_{2}$ boil-off at $\sim$240~K. The pre-cooled sample-cell assembly was then placed into the Faraday cage, the electrical connections were soldered together, and the assembly was returned to the cryostat.

The subsequent cooling and heating routines were run with a LabVIEW data acquisition program~\cite{neumeier2008} on a computer interfaced with the temperature controller and ammeter. The program measured the current, temperature, and time every $\sim$1.5 seconds while simultaneously controlling the rate of temperature change. It is also capable of maintaining a constant temperature in the cryostat to within 0.001~K and controlling the cooling/warming rates to within 0.005~K/min. However, because the cooling rate depended on the cooling power of LN$_{2}$, it decreased as \(T\) approached the boiling point of nitrogen, \(\approx 77\)~K. For \textit{slow-cooling} (\(|dT/dt|<1\)~K/min), the rates were controlled with the heater and were constant until the temperature was reached at which the heater was no longer required to slow the rate. For \textit{fast-cooling} (\(|dT/dt|>1\)~K/min), control came exclusively from the variation~\cite{neumeier2008} of the cryostat's cooling power. Curves of typical cooling rates used in this work can be seen in the insets of Figures~\ref{tsd}(a) and \ref{tsd}(b).

\section{Results}
\label{results}

\subsection{TSD current}
\label{TSDresults}

\begin{figure}[t!]
		\includegraphics[width=0.875\textwidth]{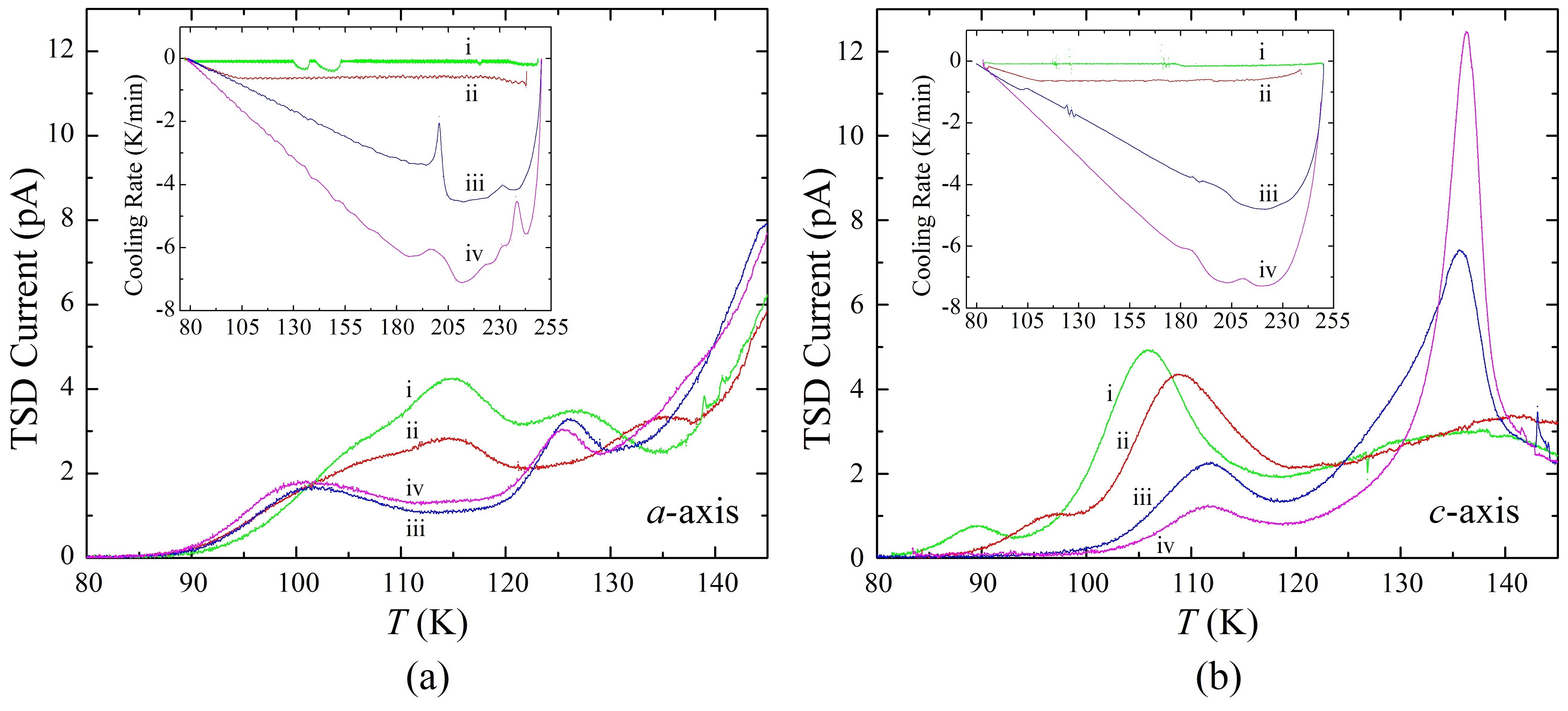}
		\caption{A plot of the TSD currents in ice measured along (a) the \(a\)-axis and (b) the \(c\)-axis. The insets show the cooling rates of each run, with the colors and labels (i, ii, iii, and iv) being the same as their counterparts in the main panels.}
		\label{tsd}
\end{figure}

In the TSD measurements, the samples were polarized for 5 minutes at 250~K with the application of 550~V to the capacitor plates, creating an electric field of 80--100~V/mm. The samples were cooled in-field at different rates as shown in the insets of Figures~\ref{tsd}(a) and \ref{tsd}(b). Once a base temperature of 80~K was reached, the electric field was removed and the sample was warmed at 0.2~K/min while the current to the capacitor plates was simultaneously measured to yield the TSD current.

The TSD currents along \(a\) and \(c\) from 80--145~K for different cooling rates are shown in Figure~\ref{tsd}(a) and \ref{tsd}(b), respectively.
Peaks in TSD currents indicate the temperatures at which relaxation processes, like the molecular reorientations in ice, occur. Each curve reveals two peaks, indicating that there exist two relaxation processes in ice in the range 80--145~K. In Figure~\ref{tsd}(a) the slow-cooled curves (i and ii) reveal a peak near 115~K that decreases in amplitude with faster cooling until it completely disappears. The fast-cooled \(a\) curves (iii and iv) are nearly identical to each other, and show peaks near 110~K and 125~K.

The TSD currents along \(c\) in Figure~\ref{tsd}(b) are more revealing. The most-slowly-cooled curve (i) shows a peak at 105~K that shifts to 112~K and decreases in amplitude as the cooling-rate increases. There is a similar effect with the peak in the slowest-cooled curves (i and ii) at 90~K that completely disappears in the fast-cooled curves (iii and iv). Another interesting feature is the development of a very sharp peak at 136~K in curves iii and iv along \(c\) which, unlike the 105~K peak, increases in amplitude upon faster cooling. This peak may be of interest as it, perhaps coincidentally, occurs at the well-known \textit{glass transition temperature} of amorphous ice,~\cite{amann-winkel2016} which is suggested~\cite{johari2002} to be caused by defect diffusion.

Also worthy of mention is the area under the TSD curves which is proportional to the charge released during warming. Along \(a\), the areas under curves ii, iii, and iv in the range 80--145~K are all equal to 85(1)\% of the area under curve i. Thus, the total charge released along \(a\) has little dependence on the cooling rate, especially for \(|dT/dt|>0.5\)~K/min. Along \(c\), however, the area under \textit{each} curve from 80--145~K depends on the rate at which it was cooled. As a percentage of the total area under curve i, the area under curves ii, iii, and iv are 98(1)\% , 87(1)\%, and 79(1)\%, respectively. Thus, slower cooling along \(c\) allows the charge to order more thoroughly and, therefore, more charge is released on warming.

\subsection{Static conductivity}
\label{SIGMAresults}

The static conductivity was determined by polarizing the sample at 250~K with the application of 550~V (\(\vec{E} \approx 90\)~V/mm) and measuring the current across the capacitor on cooling at -0.1~K/min. The current measured was the conduction current \(I_{\sigma}\) in Equation~\ref{eq:conductivity}, from which the static conductivity was calculated using \(V=550\)~V, \(A=318(1)\)~mm$^{2}$, and \(h=6.97(5)\)~mm and \(5.50(5)\)~mm for the samples with \(\vec{E}\) parallel to \(a\) and \(c\), respectively.

\begin{figure}[t!]
    \includegraphics[width=0.875\textwidth]{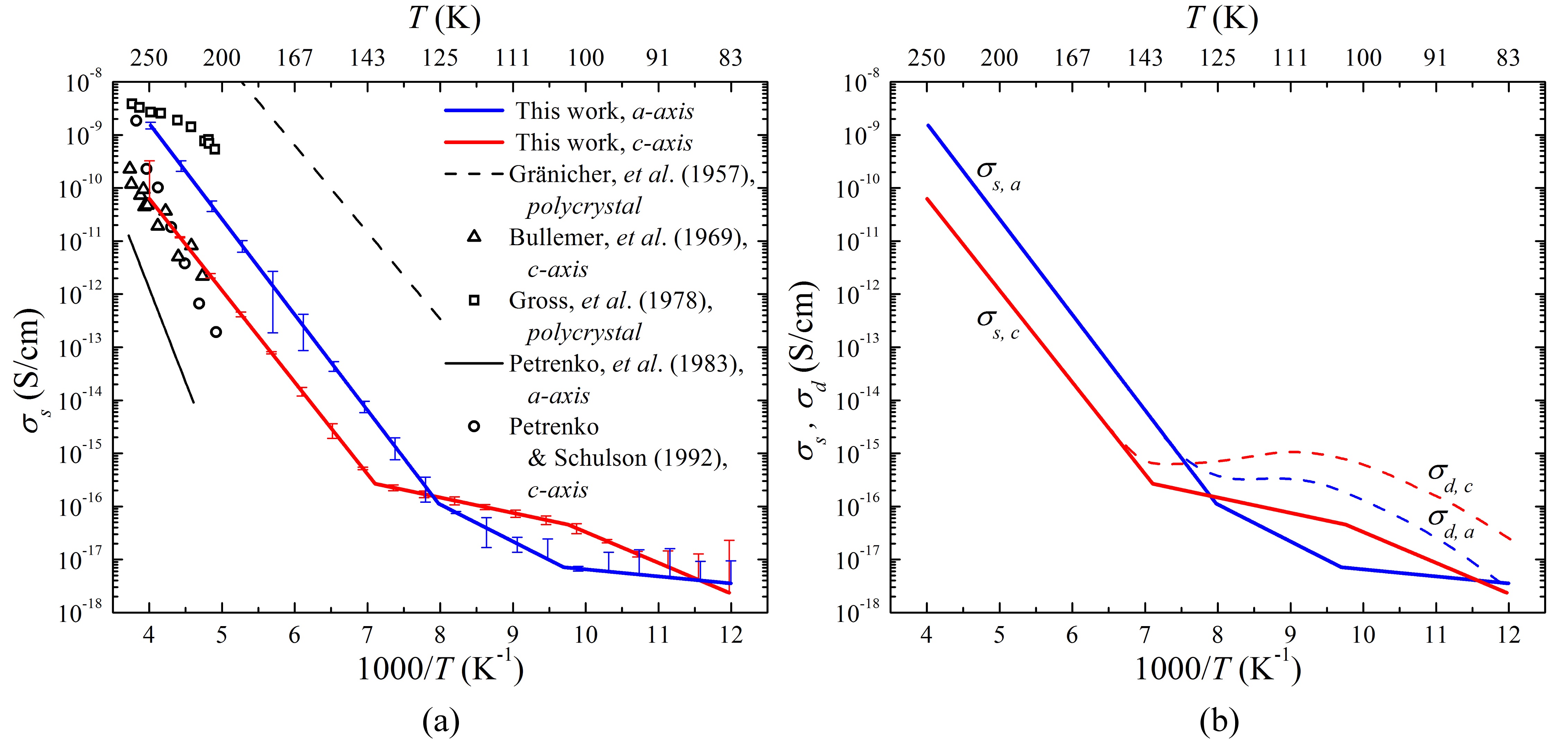}
		\caption{A log-plot of (a) the static electrical conductivity of ice \(\sigma_{s}\) versus \(1000/T\) as measured along \(a\) and \(c\), alongside published data~\cite{granicher1957, bullemer1969, gross1978, petrenko1983, petrenko1992}, and (b) \(\sigma_{s}\) alongside the predicted dielectric contribution to the conductivity \(\sigma_{d}\). The subscripts \(a\) and \(c\) in the curve labels of (b) indicate the axis along which \(\sigma_{s}\) was measured and \(\sigma_{d}\) was calculated.}
		\label{sigma}
\end{figure}

A log plot of \(\sigma_{s}\) versus \(1000/T\) measured along \(a\) and \(c\) is shown in Figure~\ref{sigma}(a),
alongside published data~\cite{granicher1957, bullemer1969, gross1978, petrenko1983, petrenko1992} for comparison. It shows \textit{three} linear regions along both axes. Error bars for the data are also included, although only 20 are shown in an effort to minimize clutter. The three \(E_{\sigma}\), found using a weighted least-square linear fit, along \(a\) are: 0.017(7)~eV from 83--103 K, 0.141(5)~eV from 103--125~K, and 0.357(1)~eV from 125--250~K. Those along \(c\) are: 0.120(4)~eV from 83--102~K, 0.059(1)~eV from 102--141~K, and 0.356(1)~eV from 141--250~K. These values reveal isotropic \(E_{\sigma}\) from 141--250~K along \(a\) and \(c\), and a clear anisotropy below 140~K. Table~\ref{Esigma}
\begin{table}[b]
\centering
\caption{Values of \(E_{\sigma}\) at 250~K from this work and the published data from Figure~\ref{sigma}(a). The letter in parentheses, if present, represents the axis (\(a\)- or \(c\)-) along which the measurement was made; if no letter is present then the sample was polycrystalline. Uncertainties are provided when available.}
\label{Esigma}
\begin{tabular}{ll}
\hline
Reference & \multicolumn{1}{c}{\(E_{\sigma}\)~(eV)} \\ \hline
This work & 0.357(1)~(\(a\)) \\
  & 0.356(1)~(\(c\)) \\
Gr{\"{a}}nicher, \textit{et~al}. & 0.325(5) \\
Bullemer, \textit{et~al}. & 0.35(2)~(\(c\)) \\
Gross, \textit{et~al}. & 0.14(1) \\
Petrenko, \textit{et~al}. & 0.70(7)~(\(a\)) \\
Petrenko $\&$ Schulson & 0.63~(\(c\)) \\ \hline
\end{tabular}
\end{table}
shows the high-temperature values of \(E_{\sigma}\) from this work and those of the authors referenced in Figure~\ref{sigma}(a). The values of \(\sigma_{s}\) at 250~K are \(1.516\times10^{-9}\)~S/cm and \(6.294\times10^{-11}\)~S/cm along \(a\) and \(c\), respectively. 

The observation of two regions of \textit{anisotropic} \(E_{\sigma}\) in Figure~\ref{sigma}(a) provides insight into the directional dependence of the conductivity. Clearly \(\sigma_{s}\) is larger along \(a\) than it is along \(c\) above 140~K, but \(E_{\sigma}\) along both axes is the same. Below 140~K, however, \(E_{\sigma}\) along \(c\) becomes smaller. This implies that conductivity along \(c\) \textit{requires less energy} than that along \(a\) in the range 100--140~K. Furthermore, along \(c\) the shape of the \(\sigma_{s}\) curve is very similar to that of the D- and L-defect conductivity in HCl-doped ice,~\cite{takei1987} supporting the general consensus~\cite{petrenko} that Debye relaxation in pure ice arises from the propagation of Bjerrum defects through molecular rotations, and not from the motion of H$_{3}$O$^{+}$ or OH$^{-}$ ions.

Another explanation is offered for the smaller slopes of the \(\sigma_{s}\) curves below 140~K as follows. The anisotropic static permittivity data of Kawada~\cite{kawada1978} from 263 to 126 K are fitted to the Curie law \(\epsilon_{a} = 23700/T\) and the Curie-Weiss law \(\epsilon_{c} = 22500/(T-46~\mathrm{K})\), respectively, where \(\epsilon_{a}\) and \(\epsilon_{c}\) are the permittivities measured along \(a\) and \(c\), respectively. It is assumed that these formulas can be extrapolated to 83~K, the lower \(T\) limit of this work. The increase in permittivity as \(T\) decreases with increasing time \(t\) causes the capacitor to receive a dielectric charging current density \(J_{d}\). Combining Equations~\ref{eq:polarization1} and \ref{eq:Idsp} yields \(J_{d}=I_{d}/A = d P/d t = \epsilon_{0} E d \epsilon/d t\) which gives a \textit{dielectric contribution to the static conductivity} \(\sigma_{d} = J_{d}/E = \epsilon_{0} d \epsilon/d t\) to the apparent conductivity. In this work, \(d \epsilon/d t = (d \epsilon/d T)(d T/d t)\), where \(d T/d t = -1/600\)~K/s, \(d \epsilon_{a}/d T = -23700/T^{2}\), and \(d \epsilon_{c}/d T = -22500/(T-46~\mathrm{K})^{2}\). These formulas explain the locations of the predicted curves, dashed lines in Figure~\ref{sigma}(b), down to 110~K. Below 110~K, the downturns of these curves result from the inability of the permittivities to follow the Curie and Curie-Weiss laws because the relaxation times, and their extrapolations from Figure~\ref{tau}, become too long. Instead, below 110~K it became necessary to step the following differential equations degree-by-degree to find the predicted permittivities for these temperatures. These equations are \(d \epsilon_{a}/d t = 23700/\tau_{Da} T\) and \(d \epsilon_{c}/d t = 22500/\tau_{Dc} (T-46~\mathrm{K})\), where \(\tau_{Da}\) and \(\tau_{Dc}\) are the relaxation times at \(T\) along \(a\) and \(c\), respectively. The predicted crossover of the conductivities along \(a\) and \(c\) occurs near the observed temperature of 127~K. Also, the predicted onsets of change in slope occur at the measured values, 125~K along \(a\) and 140~K along \(c\).  Finally, the shapes of both predicted curves are similar to the measured shape for \(\sigma_{s}\) along \(c\), but the predicted curves are well above the measured curves below 120~K. This could be a result of the extrapolation below 125~K of Kawada's Curie and Curie-Weiss law expressions for static permittivity.

Under consideration now are the extrapolations to infinite temperature of the data in Figure~\ref{sigma}(a), in terms of the discussion of intrinsic and extrinsic conductivity in Section~\ref{ssec:TSDtheory}. There, the \textit{intrinsic infinite-temperature conductivity} \(\sigma_{0,\mathrm{int}}\) was approximated to be \(2 \times 10^{3}\)~S/cm. In Figure~\ref{sigma}(a), only the extrapolations of Petrenko, \textit{et~al}.~\cite{petrenko1983} (\(a\)-axis) and Petrenko and Schulson~\cite{petrenko1992} (\(c\)-axis) converge to that value as \(T \rightarrow \infty\). Furthermore, their curves have the highest activation energies, exceeding the others by 2--6$\%$. Therefore, it is likely~\cite{petrenko} that they measured the intrinsic conductivities along these two axes, and all of the other data in Figure~\ref{sigma}(a) (including that of this work) are related to extrinsic conductivity. The value of \(\sigma_{s}\) along \(c\) determined in this work and by Bullemer~\cite{bullemer1969} have the same activation energy to within experimental uncertainty and a value of \(\sigma_{0}\) near 10$^{-3}$~S/cm. This is a factor \(f = 5 \times 10^{-7}\) times lower than \(\sigma_{0,\mathrm{int}}\), implying that the unintentional impurity fractional concentration of the samples had this \(f\) value.

\subsection{Relaxation time}
\label{TAUresults}

The dielectric relaxation time in ice was determined as follows. The samples were cooled from 250~K to 145~K at 1~K/min. Once the system reached thermal equilibrium, a 220~V step-voltage was applied across the dielectric cell and the isothermal current was measured as a function of time until it reached a constant (\(I_{\sigma}\)). The sample was then depolarized for the same length of time by flipping switch S in Figure~\ref{circuit_cell}(a) to the up position and grounding the capacitor plates. Finally, the samples were cooled by 2 or 3~K at -1~K/min and the process was repeated until the last measurement was made at 102~K. The measurement times ranged from 20~min at 145~K to 20~hr at 102~K. The currents were fitted with an exponential least-squares fit, in the form of Equation~\ref{eq:current}, which determined the value of \(\tau_{D}\). A measurement count of one every 1.5~s was determined by the time required of the ammeter to provide a repeated average of 20 readings. This set a lower limit in the determinability of \(\tau_{D}\) to 0.75~s; any lower and the sample would be almost completely polarized before the first measurement was made. An upper limit was set by the patience of the experimentalist.

\begin{figure}[t]
    \includegraphics[width=0.4375\textwidth]{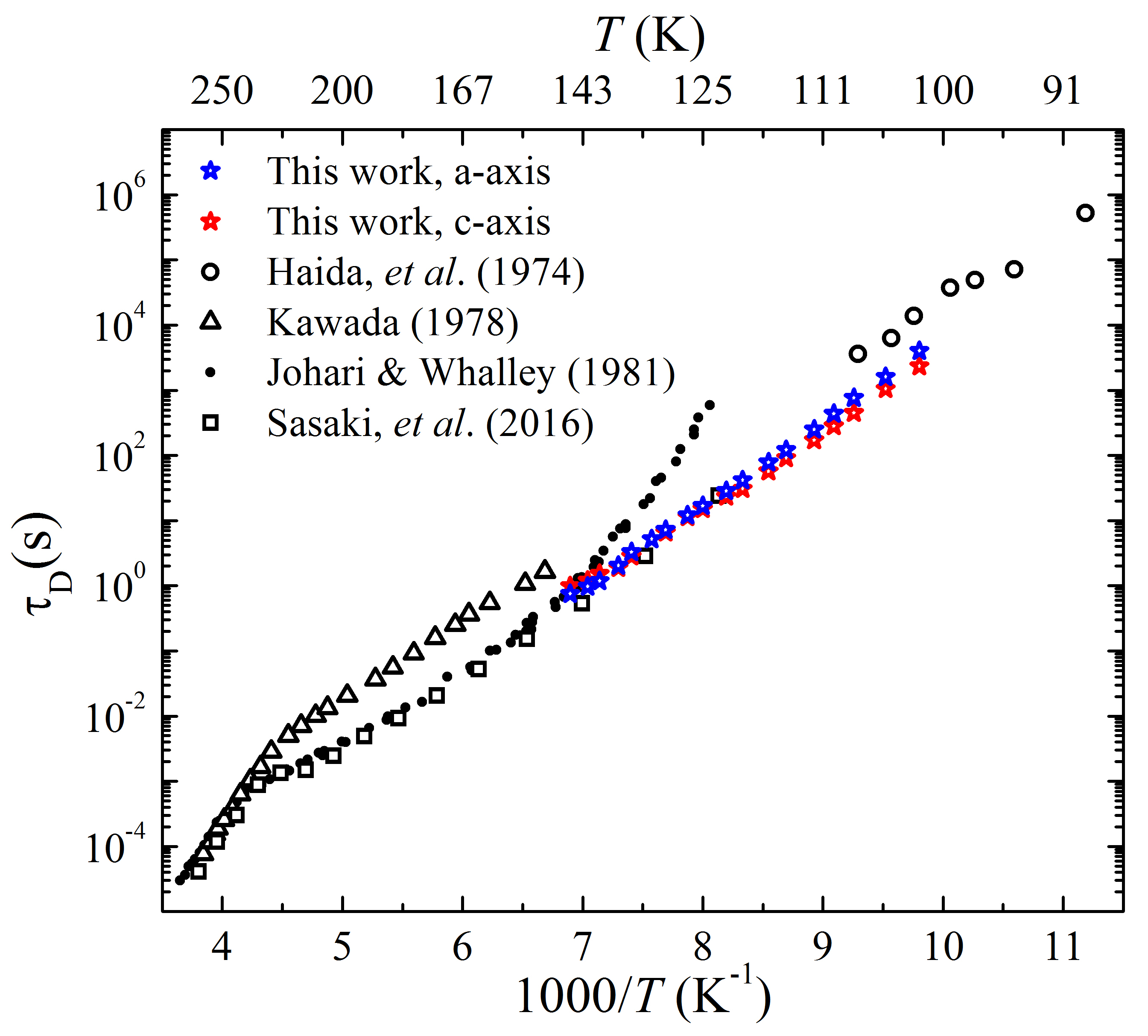}
		\caption{A plot of \(\tau_{D}\) of pure ice versus 1000/T measured along \(a\) and \(c\). The linear relationships reveal activation energies of  0.256(2)~eV and 0.233(3)~eV for \(a\) and \(c\), respectively. Published data are shown for comparison.}
		\label{tau}
\end{figure}

A log plot of \(\tau_{D}\) versus \(1000/T\) measured along \(a\) and \(c\) is shown in Figure~\ref{tau}
alongside published data.~\cite{haida1974, kawada1978, johari1981, sasaki2016} In the past, measurements of \(\tau_{D}\) were made either with polycrystals or were found to be isotropic along \(a\) and \(c\). In this work, however, anisotropy is observed which reveals \(\tau_{D}\) along \(a\) to be 1.7 times longer than that along \(c\) at 102~K. Above 135(2)~K this relationship is reversed, with \(\tau_{D}\) along \(c\) becoming longer than that along \(a\) as \(T\) increases. The relaxation times are also listed in Table~\ref{times} 
\begin{table}[t]
\centering
\caption{The dielectric (Debye) relaxation times measured along \(a\) and \(c\) in the temperature range 102--145~K. The value in parentheses is the error in the last digit.}
\label{times}
\begin{tabular}{lll}
\hline
\rule{0pt}{9pt}\vspace{1pt}\(T\)~(K) & \multicolumn{2}{c}{\(\tau_{D}\)~(s)} \\ \cline{2-3}
\rule{0pt}{11pt}\vspace{1pt} & \multicolumn{1}{c}{\(\vec{E} \parallel a\)} & \multicolumn{1}{c}{\(\vec{E} \parallel c\)} \\ \hline
\rule{0pt}{9pt}{145} & 0.7436(4) & 0.989(9) \\ 
{142} & 0.965(1) & 1.188(10) \\
{140} & 1.171(2) & 1.521(3) \\ 
{137} & 2.029(5) & 1.884(12) \\ 
{135} & 3.283(5) & 2.806(7) \\ 
{132} & 5.20(1) & 5.17(2) \\ 
{130} & 7.23(2) & 6.62(3) \\ 
{127} & 11.95(3) & 11.34(5) \\ 
{125} & 16.75(9) & 15.10(6) \\ 
{122} & 28.6(2) & 23.3(1) \\ 
{120} & 41.5(3) & 30.8(3) \\ 
{118} & 77.8(6) & 57.2(4) \\ 
{115} & 121(1) & 91.3(7) \\ 
{112} & 251(2) & 171(1) \\ 
{110} & 435(3) & 281(2) \\ 
{107} & 765(4) & 450(3) \\ 
{105} & 1608(7) & 1049(4) \\ 
\vspace{1pt}{102} & 3997(10) & 2340(15) \\ \hline
\end{tabular}
\end{table}
along with their uncertainties. The linearity predicted by Equation~\ref{eq:relaxation} allowed for the use of a linear least-squares fit of ln(\(\tau_{D}\)) versus 1/\(T\) to determine the value of \(E_{\tau}\) to be 0.256(2)~eV and 0.233(3)~eV along \(a\) and \(c\), respectively. The values of \(E_{\tau}\) reported in the published work for the temperature range 150--230~K vary between 0.195--0.234~eV, in good agreement with the values presented here.

Like the analysis of \(\sigma_{s}\) in Section~\ref{SIGMAresults}, two regimes can be considered here for the analysis of \(\tau_{D}\); relaxation by intrinsic and extrinsic carriers. In the case of \(\tau_{D}\), the carriers are expected to be Bjerrum L and D defects. For intrinsic carriers, \(\tau_{0} \approx 1/\nu \approx h/k_{B}T\). This gives a weak \(1/T\) dependence to \(\tau_{D}\) in Equation~\ref{eq:relaxation}  to the dominant exponential dependence.  Also, as seen in Figure~\ref{tau}, there are two temperature regimes with quite different activation energies above and below $\sim$230~K. These can be attributed to relaxation by intrinsic carriers above 230~K and extrinsic carriers below 230~K.

\section{Discussion}
\label{discuss}

The results of this work suggest two postulates regarding the mechanisms of dielectric relaxation dynamics in ice  below 140~K: (1) Values of the activation energies associated with relaxation in ice, and the agreement between \(\sigma_{s}\) and \(\sigma_{d}\) below 140~K, suggest those processes are dominated by the motion of Bjerrum defects. (2) Anisotropy in the data indicates that molecular reorientations in ice are energetically favored along \(c\) between 100--140~K.

Through conductivity measurements, the activation energy of L-defect mobility in HCl-doped ice was determined by Takei and Maeno~\cite{takei1987} to be 0.190(17)~eV. Similarly, Chamberlain and Fletcher~\cite{chamberlain1971} measured thermally stimulated \textit{polarization} currents in HF-doped ice and determined the activation energy for liberation of an L-defect to be 0.12(6)~eV. The agreement between these values and those of the intermediate \(E_{\sigma}\) determined in this work, 0.141~eV along \(a\) and 0.120~eV along \(c\), suggests L-defects dominate in the conductivity of pure ice below 140~K. Furthermore, in an effort to probe the partial ordering of pure ice at around 100~K, Haida, \textit{et~al}.~\cite{haida1974} determined calorimetrically (i.e. without dielectric stimulation) relaxation times associated with the proton configurational enthalpy still to be relaxed as their sample came to thermal equilibrium after a change in temperature. From these, they calculated an activation enthalpy of 0.228(40)~eV, which agrees with the \(E_{\tau}\) determined in this work, 0.256(2)~eV and 0.233(3)~eV along \(a\) and \(c\), respectively. Their relaxation times extrapolate directly into those of Kawada,~\cite{kawada1978} who determined \(\tau_{D}\) dielectrically. This reveals a common origin shared between the dielectric relaxation dynamics and the partial proton-ordering in ice near 100~K.~\cite{suga1997, yen2015.1}

Further support for the assignment of Bjerrum defects as the dominant dielectric relaxation mechanism in ice comes from the work of Schmidt.~\cite{schmidt1973} Using quadrupole perturbed deuteron NMR, the mixing time for deuterons in c-axis bonds was shown to move to the oblique bonds, and vice versa, through the propagation of Bjerrum defects. Because diffusion of an ionic defect merely moves a hydrogen along a bond, the effect does not not appreciably change the electric field gradient at the deuteron site in NMR. Therefore, Schmidt's \(\tau_{m}\) results were not sensitive to ionic diffusion. That his values of \(\tau_{m}\) extrapolate into the \(\tau_{D}\) from this work below 200~K suggests that they share the same relaxation process.

Because the relaxation process in pure (or weakly doped) ice is dominated by L-defects,~\cite{petrenko, jaccard1959} it is postulated~\cite{zaretskii1987} that peaks in the TSD current near 100~K arise from the motion of Bjerrum-defects via molecular reorientations. It is also suggested~\cite{johari1975} that the shift in temperature of the current peak near 100~K is a result of the disordered proton configuration becoming frozen-in at higher temperatures with faster warming rates. Using these ideas, the TSD currents of this work can be interpreted as follows. As ice Ih cools below 120~K, it begins to transition to the proton-ordered equilibrium state,~\cite{tajima1982} XI, through molecular rotations with a preference along \(c\). This process is halted, however, near 100~K because the dielectric relaxation time becomes so long that it effectively freezes-in the residual disorder~\cite{pauling1935} and prevents the complete transition. This behavior is well known~\cite{suga1997} but, until this work, anisotropy in this relaxation process had not been observed in pure ice. As an aside, it follows that the increase in amplitude of the 105--112~K peak in Figure~\ref{tsd}(b) indicates more proton ordering on slower cooling. Conversely, the decrease in amplitude of the 136~K peak indicates more ordering on \textit{faster} cooling.

Anisotropy has been measured in TSD experiments on KOH-doped ice by Jackson \& Whitworth,~\cite{jackson1995} with the observation of a large peak at the ferroelectric ice Ih$\rightarrow$ice XI transition temperature, 72~K, along \(c\) and a relatively short peak along \(a\). Their results indicate an anisotropic order-disorder transition whereupon molecular reorientations bring about ferroelectric order along \(c\). They suggest complete proton order along \(c\) and disorder (or alternating layers of order~\cite{rundle1954}) along \(a\). It appears that the anisotropy observed in this work is related to the aforementioned transition, but with \textit{partial} proton order along \(c\) near 100~K (as discussed in Section~\ref{TSDresults}).

The results of the conductivity measurements from this work reveal anisotropy in \(E_{\sigma}\) below 140~K. In the intermediate region from 102--140~K, one notices a smaller \(E_{\sigma}\) along \(c\) than one does along \(a\). It is also within this region, near 125~K, that the conductivity along \(c\) becomes higher than that along \(a\). Similarly, the \(\tau_{D}\) data reveals anisotropy at low temperatures, with \(E_{\tau}\) along \(a\) being larger than that along \(c\) and \(\tau_{D}\) along \(a\) being nearly twice that along \(c\) at 102~K. These \textit{independent} observations suggest that as the sample cools, the mobility of L-defects becomes less resistive along \(c\) than it is along \(a\). In other words, L-defects are energetically favored to propagate along \(c\) below 140~K.

\section{Conclusion}
\label{conclusion}

Variations of the polarization and static conductivity in pure single crystal ice with time and temperature have been measured. Values of the activation energies of conduction and dielectric relaxation suggest that polarization and relaxation in ice below 140~K are dominated by molecular rotations through propagating Bjerrum defects. Anisotropy in the TSD around 100~K, the static conductivity from 83--140~K, and the dielectric relaxation time from 102--145~K reveals that dielectric relaxation below 140~K is energetically favored along the crystallographic \(c\)-axis. It is proposed that, as ice Ih begins to transition to the proton-ordered state, ice XI, on cooling below 140~K, near 100~K the protons become more ordered along the \(c\)-axis than those along the \(a\) axis until the residual proton disorder becomes frozen-in and prevents the transition to the completely ordered state.


%
%

%

\begin{acknowledgments}
This work was funded by the National Science Foundation through NSF Award DMR-1204146. This manuscript is dedicated to the loving memory of Dr. Willis J. Buckingham.
\end{acknowledgments}

\bibliography{ice}
\bibliographystyle{aipnum4-1}

\end{document}